\documentclass[twocolumn,aps,pra,showpacs,groupedaddress,superscriptaddress,nofootinbib]{revtex4-1}

\usepackage[pdftex,linkcolor=blue,citecolor=blue,urlcolor=blue,colorlinks]{hyperref}
\usepackage[dvipsnames]{xcolor}
\usepackage{graphicx,epsfig}
\usepackage{braket}
\usepackage{color}
\usepackage{cancel}
\usepackage{inputenc}
\usepackage{amsmath}
\usepackage{amsfonts}
\usepackage{fancyhdr}
\usepackage{ulem}

\begin{document}
\title{Second-order topological corner states with ultracold atoms carrying orbital angular momentum in optical lattices}
\author{G. Pelegr\'i}
\affiliation{Departament de F\'isica, Universitat Aut\`onoma de Barcelona, E-08193 Bellaterra, Spain.}
\author{A. M. Marques}
\affiliation{Department of Physics and I3N, University of Aveiro, 3810-193 Aveiro, Portugal.}
\author{V. Ahufinger}
\affiliation{Departament de F\'isica, Universitat Aut\`onoma de Barcelona, E-08193 Bellaterra, Spain.}
\author{J. Mompart}
\affiliation{Departament de F\'isica, Universitat Aut\`onoma de Barcelona, E-08193 Bellaterra, Spain.}
\author{R. G. Dias}
\affiliation{Department of Physics and I3N, University of Aveiro, 3810-193 Aveiro, Portugal.}

\begin{abstract}
We propose a realization of a two-dimensional higher-order topological insulator with ultracold atoms loaded into orbital angular momentum (OAM) states of an optical lattice. The symmetries of the OAM states induce relative phases in the tunneling amplitudes that allow to describe the system in terms of two decoupled lattice models. Each of these models displays one-dimensional edge states and zero-dimensional corner states that are correlated with the topological properties of the bulk. We show that the topologically non-trivial regime can be explored in a wide range of experimentally feasible values of the parameters of the physical system. Furthermore, we propose an alternative way to characterize the second-order topological corner states based on the computation of the Zak's phases of the bands of first-order edge states.   
\end{abstract}
\maketitle


\section{Introduction}
Over the last decades, the study of topological insulators has become one of the most active fields in condensed matter physics \cite{reviewTopIns1,reviewTopIns2}. In these materials, the bulk-boundary correspondence establishes a relation between the topological properties of the insulating bulk and the presence of robust states at the boundaries of a finite system. Traditionally, this bulk-boundary correspondence has been considered in first-order $D$-dimensional topological insulators, where non-trivial bulk topological indices yield $(D-1)$-dimensional boundary states. In recent seminal works \cite{bernevigScience,bernevigPRB}, this concept has been extended to higher-order topological insulators (HOTIs), which display boundary modes localized in $D-n$ dimensions, with $n\geq 2$. Since their discovery, HOTIs have attracted a lot of theoretical interest \cite{bernevigScience,bernevigPRB,Theory2,Theory3,PhotonicQuadrupole,CornerWinding,TopInvariants,
BulkBoundary,AnomalousHOTI,TopInsScience,HOTISemiMetal,HOTIExact,HOTIInvSym,EntanglementHOTI,
QuadrupSemiMetals,MajoranaCorner} and have been experimentally demonstrated in several physical platforms such as metamaterials \cite{ExpPhonon, ExpAcousticKagome}, microwave \cite{ExpMicrowave}, topolectrical \cite{ExpTopEl} and LC \cite{ExpLCcircuit} circuits or solid state bismuth samples \cite{ExpBismuth}.

In this paper, we propose a scheme to realize a two-dimensional HOTI with zero-energy corner modes using ultracold atoms in optical lattices. These systems have proven to be a very versatile platform to create a variety of topological phases of matter \cite{reviewColdTop1,reviewColdTop2} in one-dimensional \cite{MeasureZak,SSHbosons1,SSHbosons2,TopologicalAnderson,SPTexpfermion} and two-dimensional \cite{HofstadterUltracold1,HofstadterUltracold2,HaldaneUltracold} settings. Our proposal is based on the use of Orbital Angular Momentum (OAM) states, which are supported by any cylindrically symmetric potential. For concreteness, we focus our discussion on ultracold atoms trapped in arrays of ring potentials, which can be implemented by a variety of techniques \cite{ring1,ring2,ring3,ring4,ring5,ring6,TAAP,TimeAveragedBECRing,TimeAveragedBECRing2,PatterningDMD,DoubleRing,ring7} and where OAM can be directly transferred to the atoms using focused light beams \cite{OptOAMAtoms}. Alternatively, OAM states can also be created in conventional optical lattices by exciting the atoms to the $p$-band \cite{HigherOrbital1,HigherOrbital2,HigherOrbital3,HigherOrbital4} or periodically modulating the lattice amplitude \cite{MugaOAM}. The distinctive advantage of OAM states is that they give rise to complex tunneling amplitudes in a natural way \cite{geometricallyinduced,EdgeLikeStates}, constituting an alternative to artificially engineered gauge fields \cite{ArtGaugeFields1,ArtGaugeFields2,ArtGaugeFields3,ArtGaugeFields4}. The relative phase between these complex tunneling amplitudes can be tuned by modifying the geometry of the system. Taking advantage of this fact, we consider a lattice in which the arrangement of the relative phases allows one to decouple the full model with two OAM states per site into two independent lattices with only one orbital per site. These lattices are just rotated versions of each other and thus share the same topological phases, giving rise to non-trivial topology in the global system. The latter is signalled by the presence of both edge states, related to weak topological properties, and zero-energy corner states, which are associated to second-order topological invariants.

The rest of the paper is organized as follows. In Sec.~\ref{physical} we describe the physical system, introduce the basis rotation that decouples the full system with two OAM orbitals per site into two independent lattices with one orbital per site and we analyze the band structure of the resulting subsystems. In Sec.~\ref{TopProp} we perform the topological characterization of the system. First, we discuss the weak topological properties that give rise to the edge states. We then move on to analyze the second-order effects, and we propose an alternative way to predict the presence of corner states by computing the Zak's phases \cite{ZakPhase} of the bands of first-order edge states. Finally, in Sec.~\ref{conclusions} we summarize the main conclusions of this work.
\section{Physical system}
\label{physical}
\begin{figure}[t!]
\centering
\includegraphics[width=0.88\linewidth]{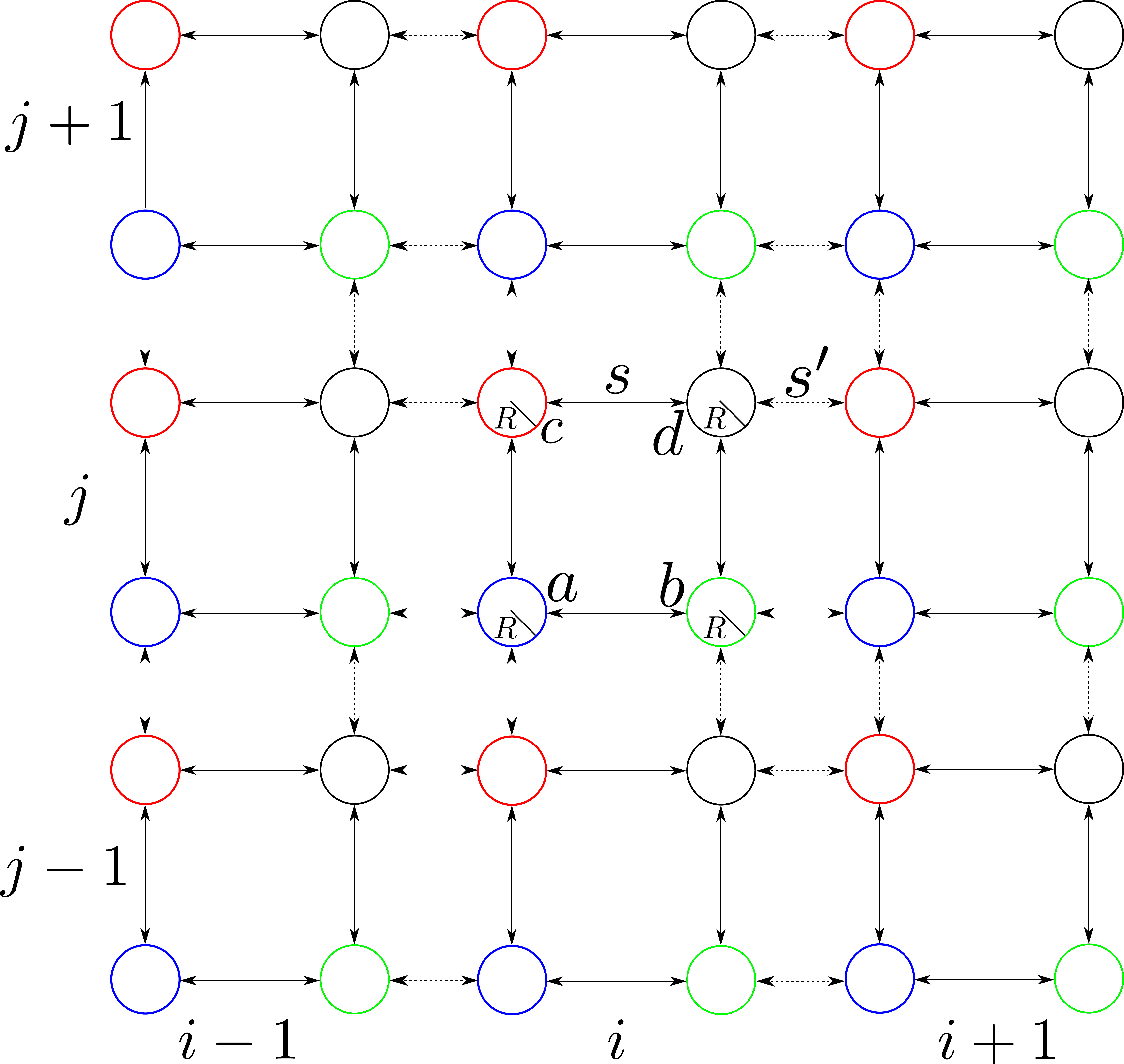}
\caption{Sketch of the two-dimensional lattice of rings of radius $R$ considered in this work. The unit cells are formed by 4 rings, named $\{a,b,c,d\}$, with intra- and inter-cell separation between their outermost parts $s$ and $s'$, respectively.}
\label{lattice}
\end{figure}
The physical system that we consider consists of a gas of non-interacting ultracold atoms of mass $m$ trapped in a two-dimensional lattice with a unit cell formed by four sites which we denote $\{a,b,c,d\}$, as depicted in Fig.~\ref{lattice}. Each of the sites is the center of a ring-shaped optical trap of radius $R$, and the intra- and inter-cell separations between the outermost parts of the rings are $s$ and $s'$, respectively. Such a lattice could be created by means of several different techniques. On the one hand, one could use time-averaged adiabatic potentials  \cite{ring4,TAAP,ring5,TimeAveragedBECRing,TimeAveragedBECRing2}, which have proven to be a versatile tool to create on-demand dynamic potential landscapes for trapping ultracold atoms. In the last few years, the possibility to use digital micro-mirror devices \cite{PatterningDMD} to create potentials with arbitrary shapes, including a double ring trap \cite{DoubleRing}, has also been demonstrated. Trapping of ultracold atoms in ring geometries has also been achieved with conical refraction \cite{ring7}. These already demonstrated approaches could be adapted to produce the two-dimensional arrangements of ring potentials that we consider in this work. Each of the ring traps that form the lattice creates a potential $V(r)=\frac{1}{2}m\omega^2(r-R)^2$, which defines a radial length scale $\sigma=\sqrt{\frac{\hbar}{m\omega}}$, where $\omega$ is the radial frequency and $\hbar$ the reduced Planck's constant. These potentials support modes with an integer OAM $l$. For concreteness, in this paper we focus on the particular case in which the atoms may occupy the two degenerate excited OAM states $l=1$ with positive or negative circulation of each ring potential of the lattice, but all our considerations could be generalized to higher OAM states in a straightforward manner. We denote the OAM $l=1$ states with positive and negative circulations as $\ket{p_{i,j},\pm}$, where $i,j$ are indices that indicate the horizontal and vertical positions of the unit cell and $p=\{a,b,c,d\}$ labels the site. The wavefunctions of these states are given by
\begin{equation}
\phi_{\pm}^{p_{i,j}}(r_{p_{i,j}},\varphi_{p_{i,j}})=\braket{\vec{r}|p_{i,j},\pm}=\psi(r_{p_{i,j}})e^{\pm i(\varphi_{p_{i,j}}-\varphi_0)},
\label{wavefunctions}
\end{equation}
where $(r_{p_{i,j}},\varphi_{p_{i,j}})$ are the polar coordinates with origin at the site $p_{i,j}$ and $\varphi_0$ is an absolute phase origin, which can be chosen arbitrarily. 

In order to derive the Hamiltonian of the system that we consider, let us summarize briefly the arguments presented in \cite{geometricallyinduced}, where the tunneling dynamics of OAM states was studied in detail. We first consider a system formed by only two of the rings that form the lattice, which we name $\lambda$ and $\rho$. The tunneling amplitudes between the four states that form the OAM $l=1$ manifold of this two-ring system are given by the following overlap integrals 
\begin{align}
&J_{\beta,n}^{\alpha,p}=e^{i(p-n)\varphi_0}\times\nonumber\\
&\times\int \left(\phi_p^{\alpha}(\varphi_0=0)\right)^* \left[-\frac{\hbar^2\nabla^2}{2m}+V(\vec{r})\right] \phi_n^{\beta}(\varphi_0=0) d\vec{r},
\label{OverlapCouplings}
\end{align}
where $V(\vec{r})$ is the total potential of the two-ring system, $\alpha,\beta=\lambda,\rho$ and $n,p=\pm$. By analysing the mirror symmetries, one realizes that there are only three independent tunneling amplitudes. We denote them as: i) $J_1(R,s)\equiv J_{\alpha,n}^{\alpha,-n}$, which corresponds to the self-coupling between the two OAM states of each trap induced by the breaking of the global cylindrical symmetry of the problem, ii) $J_2(R,s)\equiv J_{\lambda,n}^{\rho,n}$, which corresponds to the cross-coupling between states in different sites with the same circulation, and iii) $J_3(R,s)\equiv J_{\lambda,n}^{\rho,-n}$, which corresponds to the cross-coupling between states in different sites with different circulations. Note that we have explicitly stated the dependence of the couplings on the radius of the rings $R$ and the separation between them $s$, which determine the absolute and relative strength of the different tunneling amplitudes \cite{diamondchain}. Combining the hermiticity of the Hamiltonian with the analysis based on the mirror symmetries, it can be shown that the integral appearing in \eqref{OverlapCouplings} is real. Therefore, the origin of azimuthal phases $\varphi_0$ induces a $e^{i(p-n)\varphi_0}$ factor in the tunneling couplings. Note that these phases can only appear in the tunneling amplitudes corresponding to an exchange of the circulation of the OAM states, i.e. $J_1(R,s)$ and $J_3(R,s)$, for which $p\neq n$. In a two-trap system, one can always take  $\varphi_0=0$ and thus all the tunneling couplings become real \cite{geometricallyinduced}. However, in a system formed by more than two traps that are not aligned such as, for instance, the unit cell of the 2D lattice depicted in Fig.~\ref{lattice}, relative phases in the tunneling amplitudes appear due to the fact that there is a relative angle between the line defining the origin of phases and at least one of the lines connecting the centers of the traps. These phases are a natural consequence of the form of the wavefunctions of the OAM states, and can be modulated by tuning the geometry of the system. In quasi-one dimensional systems, they can be used to engineer lattices with topological edge states and Aharonov-Bohm caging \cite{diamondchain}. Choosing the origin of phases along the line that unites sites $a_{i,j}\leftrightarrow b_{i,j}$ in the lattice of Fig.~\ref{lattice}, along the perpendicular direction $a_{i,j}\leftrightarrow c_{i,j}$ $\varphi_0=\pi/2$, and therefore the couplings acquire a relative $\pi$ phase. Moreover, destructive interference between the contribution of neighbouring sites causes the self-coupling terms $J_1$ to vanish \cite{geometricallyinduced}. Thus, the Hamiltonian of the noninteracting system reads
\begin{align}
\hat{H}&=J_2\sum_{i,j}\sum_{\alpha=\pm}
\hat{a}_{\alpha}^{i,j\dagger}(\hat{b}_{\alpha}^{i,j}+\hat{c}_{\alpha}^{i,j})+
\hat{d}_{\alpha}^{i,j\dagger}(\hat{b}_{\alpha}^{i,j}+\hat{c}_{\alpha}^{i,j})
\nonumber\\
&+J_2'\sum_{i,j}\sum_{\alpha=\pm}
\hat{a}_{\alpha}^{i,j\dagger}(\hat{b}_{\alpha}^{i-1,j}+\hat{c}_{\alpha}^{i,j-1})+
\hat{d}_{\alpha}^{i,j\dagger}(\hat{b}_{\alpha}^{i,j+1}+\hat{c}_{\alpha}^{i+1,j})
\nonumber\\
&+J_3\sum_{i,j}\sum_{\alpha=\pm}
\hat{a}_{\alpha}^{i,j\dagger}(\hat{b}_{-\alpha}^{i,j}-\hat{c}_{-\alpha}^{i,j})+
\hat{d}_{\alpha}^{i,j\dagger}(-\hat{b}_{-\alpha}^{i,j}+\hat{c}_{-\alpha}^{i,j})
\nonumber\\
&+J_3'\sum_{i,j}\sum_{\alpha=\pm}
\hat{a}_{\alpha}^{i,j\dagger}(\hat{b}_{-\alpha}^{i-1,j}-\hat{c}_{\alpha}^{i,j-1})+
\hat{d}_{\alpha}^{i,j\dagger}(-\hat{b}_{-\alpha}^{i,j+1}+\hat{c}_{\alpha}^{i+1,j})
\nonumber\\
&+\text{H.c.},
\label{Ham_2DSSH_OAM}
\end{align}
where we have defined $J_{2(3)}\equiv J_{2(3)}(R,s)$, $J_{2(3)}'\equiv J_{2(3)}(R,s')$ and the annihilation operators $\hat{p}_{\pm}^{i,j}$ associated to the states $\ket{p_{i,j},\pm}$. The single-particle properties derived from the Hamiltonian \eqref{Ham_2DSSH_OAM_Asym} are independent of the quantum statistics. However, in some cases we will compute quantities that involve occupation by a single atom of consecutive quantum levels. In those instances, we will assume a spinless fermionic species, because non-interacting bosons would accumulate into the lowest-energy state.

\begin{figure}[t!]
\centering
\includegraphics[width=\linewidth]{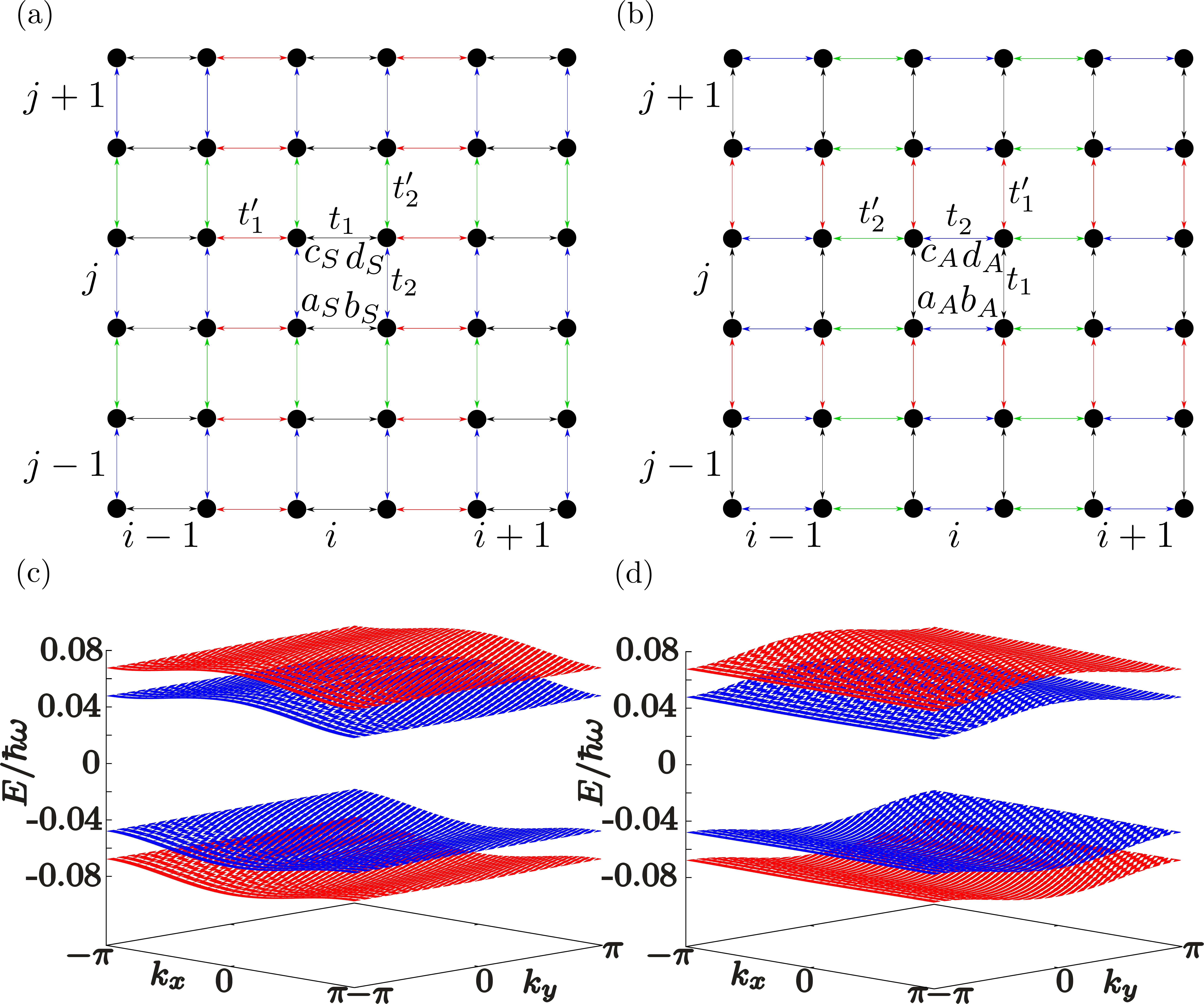}
\caption{Sketches of the two-dimensional lattices of (a) symmetric and (b) antisymmetric combinations of OAM orbitals, which are described respectively by the Hamiltonians \eqref{Ham_2DSSH_OAM_Sym} and \eqref{Ham_2DSSH_OAM_Asym}. Band structures of the (c) symmetric and (d) antisymmetric lattices. The parameters of the physical lattice are $R=2.5\sigma$, $s=4\sigma$ and $s'=2\sigma$, for which the coupling parameters of the symmetric and antisymmetric lattices are $t_1/t_1'=0.09$, $t_2/t_2'=0.03$, $t_2'/t_1'=-0.16$.}
\label{SymAsymLattices}
\end{figure}

In order to simplify the treatment of the model, we consider a new basis formed by the symmetric and anti-symmetric combinations of OAM states with opposite circulation at each site, the density profiles of which resemble those of $p_x$ and $p_y$ orbitals, respectively.   
\begin{align}
&\ket{p_{i,j},S}=\frac{1}{\sqrt{2}}(\ket{p_{i,j},+}+\ket{p_{i,j},-}),\\
&\ket{p_{i,j},A}=\frac{1}{\sqrt{2}}(\ket{p_{i,j},+}-\ket{p_{i,j},-}).
\end{align}
In this rotated basis, the lattice with two OAM orbitals per site described by the Hamiltonian \eqref{Ham_2DSSH_OAM} gets decoupled into two independent lattices with only one symmetric or anti-symmetric orbital per site that are related to each other by a $C_4$ rotation  
\begin{align}
\hat{H}&=\hat{H}_S+\hat{H}_A,\\
\hat{H}_S&=\sum_{i,j}
t_1(\hat{a}_{S}^{i,j\dagger}\hat{b}_{S}^{i,j}+
\hat{c}_{S}^{i,j\dagger}\hat{d}_{S}^{i,j})+t_1'(\hat{b}_{S}^{i,j\dagger}\hat{a}_{S}^{i+1,j}+
\hat{d}_{S}^{i,j\dagger}\hat{c}_{S}^{i+1,j})
\nonumber\\
&+\sum_{i,j}t_2(\hat{a}_{S}^{i,j\dagger}\hat{c}_{S}^{i,j}+
\hat{b}_{S}^{i,j\dagger}\hat{d}_{S}^{i,j})+
t_2'(\hat{c}_{S}^{i,j\dagger}\hat{a}_{S}^{i,j+1}+
\hat{d}_{S}^{i,j\dagger}\hat{b}_{S}^{i,j+1})
\nonumber\\
&+\text{H.c.},
\label{Ham_2DSSH_OAM_Sym}
\\
\hat{H}_A&=\sum_{i,j}
t_2(\hat{a}_{A}^{i,j\dagger}\hat{b}_{A}^{i,j}+
\hat{c}_{A}^{i,j\dagger}\hat{d}_{A}^{i,j})+t_2'(\hat{b}_{A}^{i,j\dagger}\hat{a}_{A}^{i+1,j}+
\hat{d}_{A}^{i,j\dagger}\hat{c}_{A}^{i+1,j})
\nonumber\\
&+\sum_{i,j}t_1(\hat{a}_{A}^{i,j\dagger}\hat{c}_{A}^{i,j}+
\hat{b}_{A}^{i,j\dagger}\hat{d}_{A}^{i,j})+
t_1'(\hat{c}_{A}^{i,j\dagger}\hat{a}_{A}^{i,j+1}+
\hat{d}_{A}^{i,j\dagger}\hat{b}_{A}^{i,j+1})
\nonumber\\
&+\text{H.c.},
\label{Ham_2DSSH_OAM_Asym}
\end{align}
where we have defined the new coupling constants $t_1\equiv J_2+J_3,t_1'\equiv J_2'+J_3',t_2\equiv J_2-J_3,t_2'\equiv J_2'-J_3'$. Both $\hat{H}_S$ and $\hat{H}_A$ possess chiral and $x$ and $y$ reflection symmetries. The lattices of symmetric and anti-symmetric orbitals described by the Hamiltonians \eqref{Ham_2DSSH_OAM_Sym} and \eqref{Ham_2DSSH_OAM_Asym} are represented in Figs.~\ref{SymAsymLattices} (a) and (b), respectively. They differ from the minimal model of a topological quadrupole insulator proposed in \cite{bernevigPRB} by the fact that the cells of the lattices are not threaded by a net flux and have distinct staggering patterns for the coupling parameters along both directions ($t_1-t_1^\prime$ and $t_2-t_2^\prime$) which, in a sense, mimics the effect of a finite magnetic flux, in what concerns the opening of the energy gap around zero energy. As such, this system can also display second-order topological corner states and quadrupole moment as we will show in the following section.

Using the $\{a,d,b,c\}$ ordering for the $k-$space basis in order to make manifest the chiral symmetry, the bulk Hamiltonians of the symmetric and antisymmetric lattices read
\begin{equation}
H_S=\begin{pmatrix}
0&0&t_1+t_1'e^{-ik_x}&t_2+t_2'e^{-ik_y}\\
0&0&t_2+t_2'e^{ik_y}&t_1+t_1'e^{ik_x}\\
t_1+t_1'e^{ik_x}&t_2+t_2'e^{-ik_y}&0&0\\
t_2+t_2'e^{ik_y}&t_1+t_1'e^{-ik_x}&0&0
\end{pmatrix}
\label{Hk_sym}
\end{equation}
\begin{equation}
H_A=\begin{pmatrix}
0&0&t_2+t_2'e^{-ik_x}&t_1+t_1'e^{-ik_y}\\
0&0&t_1+t_1'e^{ik_y}&t_2+t_2'e^{ik_x}\\
t_2+t_2'e^{ik_x}&t_1+t_1'e^{-ik_y}&0&0\\
t_1+t_1'e^{ik_y}&t_2+t_2'e^{-ik_x}&0&0
\end{pmatrix}
\label{Hk_asym}
\end{equation}
and their corresponding energy bands are given by
\begin{subequations}
\begin{align}
E_S^{1}&=-E_S^4=-T_1(k_x)-T_2(k_y),\\
E_S^{2}&=-E_S^3=-T_1(k_x)+T_2(k_y),
\end{align}
\label{bands_sym}
\end{subequations}
\begin{subequations}
\begin{align}
E_A^{1}&=-E_A^4=-T_2(k_x)-T_1(k_y),\\
E_A^{2}&=-E_A^3=-T_2(k_x)+T_1(k_y),
\end{align}
\label{bands_asym}
\end{subequations}
where $T_q(k_\mu)=\sqrt{t_q^2+t_q'^2+2t_qt_q'\cos(k_\mu)}$. The band structures \eqref{bands_sym} and \eqref{bands_asym} are gapped at zero energy if the couplings fulfill either the relation ${|t_1-t_1'|>|t_2+t_2'|}$ or ${|t_2-t_2'|>|t_1+t_1'|}$. Owing to the dependence of the couplings of the original model \eqref{Ham_2DSSH_OAM} $J_{2(3)}^{(')}$ on the parameters of the system \cite{diamondchain}, these conditions are fulfilled for a wide range of experimentally reasonable values of $R$, $s$ and $s'$. This is exemplified in Figs. \ref{SymAsymLattices} (c) and (d), where the gapped band structures of the symmetric and anti-symmetric lattices that are obtained for the coupling parameters corresponding to rings of radius $R=2.5\sigma$ with intra- and inter-cell separations $s=4\sigma$ and $s'=2\sigma$ are shown. In the next section, we discuss the topological properties of the model and show how they manifest themselves through the presence of edge and corner states in finite systems. Since the lattices of symmetric and anti-symmetric orbitals described by the Hamiltonians \eqref{Ham_2DSSH_OAM_Sym} and \eqref{Ham_2DSSH_OAM_Asym} are related by a $C_4$ rotation, it is enough to analyze only one of them in order to characterize the full model with two OAM orbitals per site. In the following, we will focus the discussion on the lattice of symmetric orbitals. Although it is not necessary to experimentally distinguish between symmetric and antisymmetric orbitals in order to observe the properties of the system that we shall discuss, we note that in some physical platforms supporting $p_x$ and $p_y$ orbitals it is possible to manipulate separately the lattices described by the models \eqref{Ham_2DSSH_OAM_Sym} and \eqref{Ham_2DSSH_OAM_Asym}. In the $p$-band of a conventional optical lattice \cite{HigherOrbital1}, this could be done by using lasers with different intensities along $x$ and $y$, in such a way that the gaps between the $s$ and $p$ bands would be different along each direction and the energies of the $p_x$ and $p_y$ orbitals would be shifted. Energy shifting and separate manipulation of the $p_x$ and $p_y$ orbitals has been demonstrated in an artificial electronic lattice \cite{pBandElectronic}, which is another physical platform where the model studied in this work could be implemented.
\begin{figure}[t!]
\centering
\includegraphics[width=\linewidth]{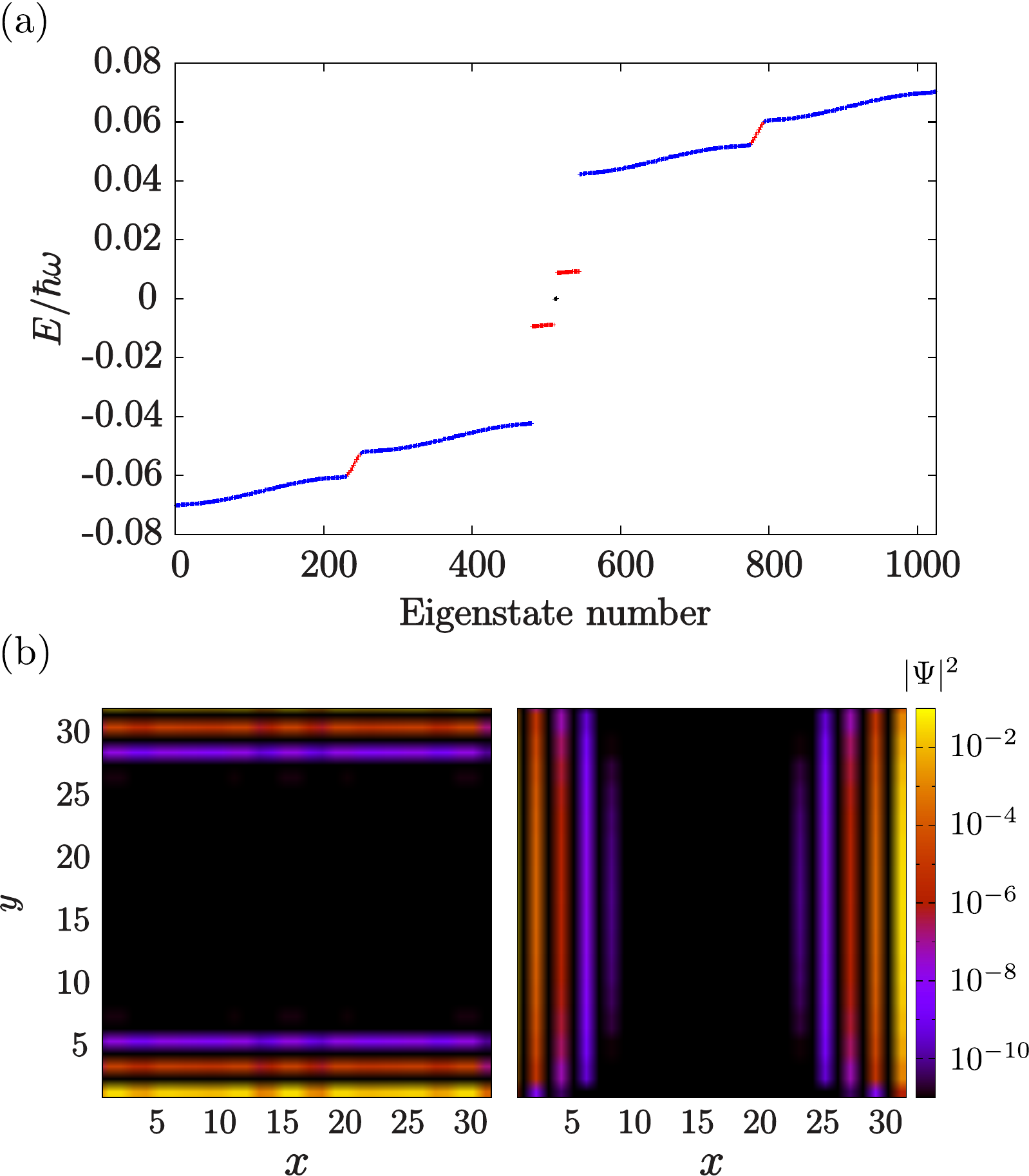}
\caption{(a) Full energy spectrum with bulk states (blue dots), horizontal and vertical edge states (red dots), and zero-energy corner states (black dots). (b) Density profile of a vertical (horizontal) edge state at the left (right). All plots correspond to a symmetric lattice of $16\times 16$ unit cells. The parameters of the physical lattice are $R=2.5\sigma$, $s=4\sigma$ and $s'=2\sigma$, for which the coupling parameters of the symmetric lattice are $t_1/t_1'=0.09$, $t_2/t_2'=0.03$, $t_2'/t_1'=-0.16$.}
\label{SpectrumStates}
\end{figure}
\section{Topological properties}
\label{TopProp}
Let us consider a lattice of symmetric orbitals as the one depicted in Fig. \ref{SymAsymLattices} (a) formed by $N_x$ and $N_y$ unit cells along the $x$ and $y$ directions respectively. In the limit of zero intra-cell couplings, $t_1=t_2=0$, corner and edge states appear naturally in this lattice. The four corner sites are completely decoupled from the rest of the system, and therefore they arise as zero-energy states in the spectrum. Moreover, the horizontal edges are composed of $N_x-2$ isolated dimers with internal coupling $t_1'$. Thus, the spectrum also has $2N_x-4$ vertical edge states, of which one half have energy $t_1'$ and the other half $-t_1'$. Similarly, the $y$ edges host $2N_y-4$ horizontal edge states, of which one half have energy $t_2'$ and the other half $-t_2'$. All of these states have a topological origin, and are therefore present in the energy spectrum beyond the limit of null intra-cell couplings. This is illustrated in Fig. \ref{SpectrumStates} (a), which shows the spectrum of a lattice of $16\times 16$ unit cells formed by rings of radius $R=2.5\sigma$ and inter- and intra-cell separations $s=4\sigma$ and $s'=2\sigma$, for which the coupling parameters fulfill the relations $|t_1'|>|t_1|,|t_2'|>|t_2|$ and are such that all the gaps are open. While the corner states remain at zero energy, the horizontal and vertical edge states (labeled according to the direction over which they decay) change their energies with respect to the $t_1=t_2=0$ limit and form dispersive bands. Examples of the density profiles of the vertical and horizontal edge states are shown in Fig. \ref{SpectrumStates} (b).

The topological mechanisms that give rise to the edge and the corner states are different. While the former can be understood in terms of the two-dimensional Zak's phase \cite{2DSSHoneorbital}, the latter are due to second-order topological effects \cite{bernevigPRB}. Thus, in the next subsections we discuss separately these two different mechanisms, and we then combine all the results to fully characterize the topological phase diagram of the model. 

\subsection{Weak topology and edge states}
The model under consideration does not constitute a Chern insulator \cite{reviewTopIns1,reviewTopIns2}, since it has a vanishing Chern number. This is a consequence of the fact that the model is invariant under both time-reversal and inversion symmetry. The former implies that the Berry curvature of each band is an odd function of $\vec{k}$, $\Omega_n(\vec{k})=-\Omega_n(-\vec{k})$, while the latter imposes that it must be an even function of $\vec{k}$, $\Omega_n(\vec{k})=\Omega_n(-\vec{k})$. In order to satisfy both constraints simultaneously the Berry curvatures must vanish everywhere in quasimomentum space, implying that the Chern number is 0 for all energy bands \cite{BerryPhase}.

According to the modern theory of polarization \cite{ModPolarization}, the edge states are related to the polarization properties of the bulk. In turn, these properties are directly related to the topology of the model, which can be characterized using the Wilson-loop approach. This formalism, which was developed in the context of solid state physics, can be directly adapted to systems of ultracold atoms in optical lattices by identifying the negative/positive charges with bright/dark peaks in the atomic density distributions. 

Let us consider the Bloch functions $\ket{u_S^i(\mathbf{k})}$, which are the eigenvectors associated to the energy bands $E_S^i$ defined in eqs. \eqref{bands_sym}. From them, we define the Wilson-loop operators along the $x$ and $y$ directions $\mathcal{W}_x(k_y)$ and $\mathcal{W}_y(k_x)$, the matrix elements of which are given by
\begin{align}
&\mathcal{W}_x^{i,j}(k_y)=\prod_{n=0}^{N-1} \braket{u_S^i(k_x+n\Delta k,ky)|u_S^j(k_x+(n+1)\Delta k,k_y)}\\
&\mathcal{W}_y^{i,j}(k_x)=\prod_{n=0}^{N-1} \braket{u_S^i(k_x,ky+n\Delta k)|u_S^j(k_x,k_y+(n+1)\Delta k)},
\end{align}
where $N$ is the number of discrete points along each of the directions in $k-$space and $\Delta k=\frac{2\pi}{N}$. The indices of the matrix elements run in the range $i,j=1,...,N_{\text{occ}}$, where $N_{\text{occ}}$ is the number of occupied bands.
\begin{figure}[t!]
\centering
\includegraphics[width=\linewidth]{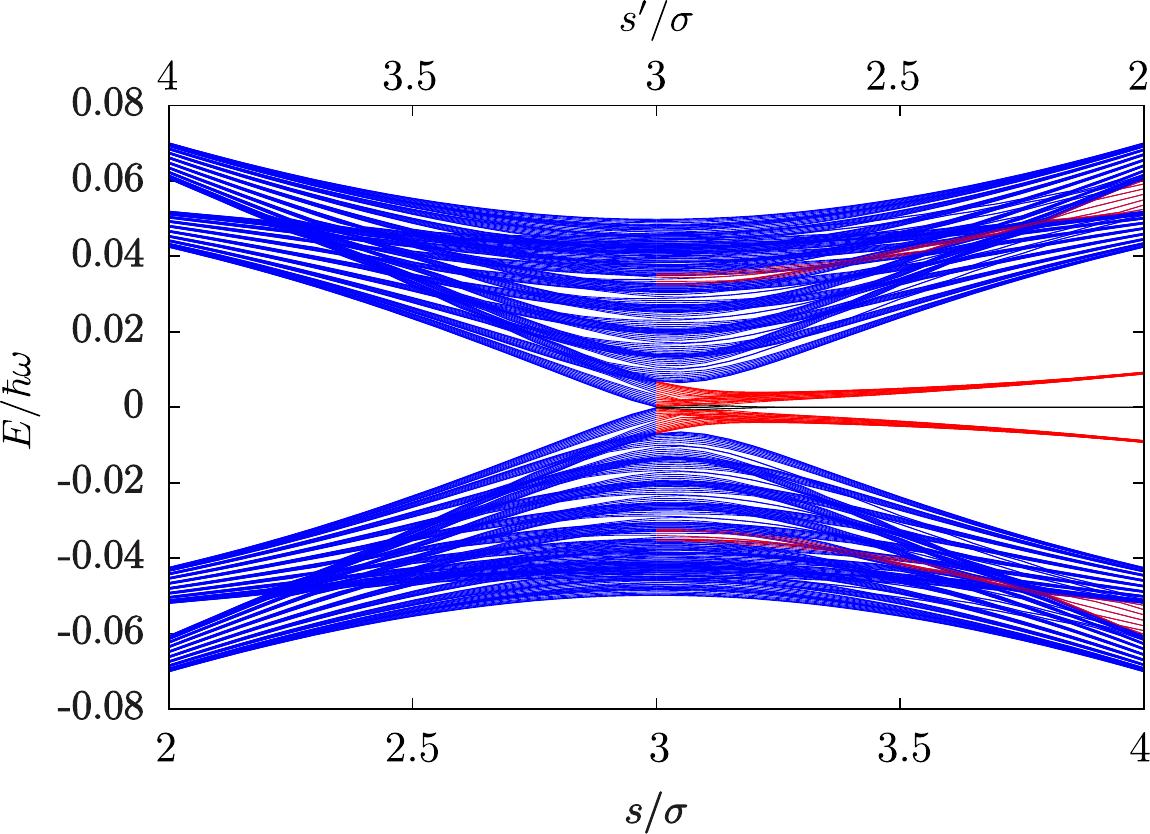}
\caption{Energy spectrum of an open symmetric lattice of $10\times 10$ unit cells formed by rings of radius $R=2.5\sigma$ as a function of the inter-ring separation. 
Blue, red and black curves correspond to bulk, edge and corner states, respectively.}
\label{spectrum_sym_distance}
\end{figure}
The symmetric lattice described by the Hamiltonian \eqref{Ham_2DSSH_OAM_Sym} has two bands below the gap centered around zero energy. Therefore, for a non-interacting spinless fermionic gas $N_{\text{occ}}=2$ at half filling. Since the Wilson-loop operators are unitary in the thermodynamic limit (where $N\to \infty$ and $\Delta k\to 0$), their eigenvalues are phases. From the Wilson-loop operators, we define the Wannier Hamiltonians \cite{bernevigPRB}
\begin{align}
&H_{\mathcal{W}_x}(k_y)=-\frac{i}{2\pi} \ln \mathcal{W}_x(k_y)\\
&H_{\mathcal{W}_y}(k_x)=-\frac{i}{2\pi} \ln \mathcal{W}_y(k_x),
\end{align}
whose eigenvalues and eigenvectors are denoted as $\nu_x^j(k_y), \nu_y^j(k_x)$ and $\ket{\nu_x^j(k_y)}, \ket{\nu_y^j(k_x)}$ ($j=1,...,N_{\text{occ}}$). The eigenvalues at each point in $k-$space are known as the Wannier centers, and the set of all the Wannier centers form the so-called Wannier bands \cite{bernevigPRB}. Finally, the $x$ and $y$ bulk polarizations can be computed from the Wannier bands as
\begin{align}
P_x=\frac{1}{2\pi}\sum_{j=1}^{N_{\text{occ}}}\int_{0}^{2\pi} dk_y\nu_x^j(k_y)\equiv \sum_{j=1}^{N_{\text{occ}}} P_x^j \text{ (mod }1)\\
P_y=\frac{1}{2\pi}\sum_{j=1}^{N_{\text{occ}}}\int_{0}^{2\pi} dk_x\nu_y^j(k_x)\equiv \sum_{j=1}^{N_{\text{occ}}} P_y^j \text{ (mod }1).
\end{align}
In the $N\rightarrow \infty$ limit, the polarizations can also be computed as $P_x^j=\frac{1}{2\pi}\gamma_x^j$, $P_y^j=\frac{1}{2\pi}\gamma_y^j$, where $\gamma_x^j$ and $\gamma_x^j$ are the two-dimensional generalizations of the Zak's phase of the band $j$,
\begin{align}
&\gamma_x^j=\frac{i}{2\pi}\int_{\text{BZ}}d\mathbf{k} \braket{u_S^j(\mathbf{k})|\partial_{k_x}|u_S^j(\mathbf{k})}\label{2DZakX}\\
&\gamma_y^j=\frac{i}{2\pi}\int_{\text{BZ}}d\mathbf{k} \braket{u_S^j(\mathbf{k})|\partial_{k_y}|u_S^j(\mathbf{k})}.
\label{2DZakY}
\end{align}
Our model has reflection symmetry in the $x$ and $y$ directions. In this situation, the 2D Zak's phases are quantized to 0 or $\pi$, and therefore the total polarizations can only be $0$ or $1/2$ for both directions.

\begin{figure}[t!]
\centering
\includegraphics[width=\linewidth]{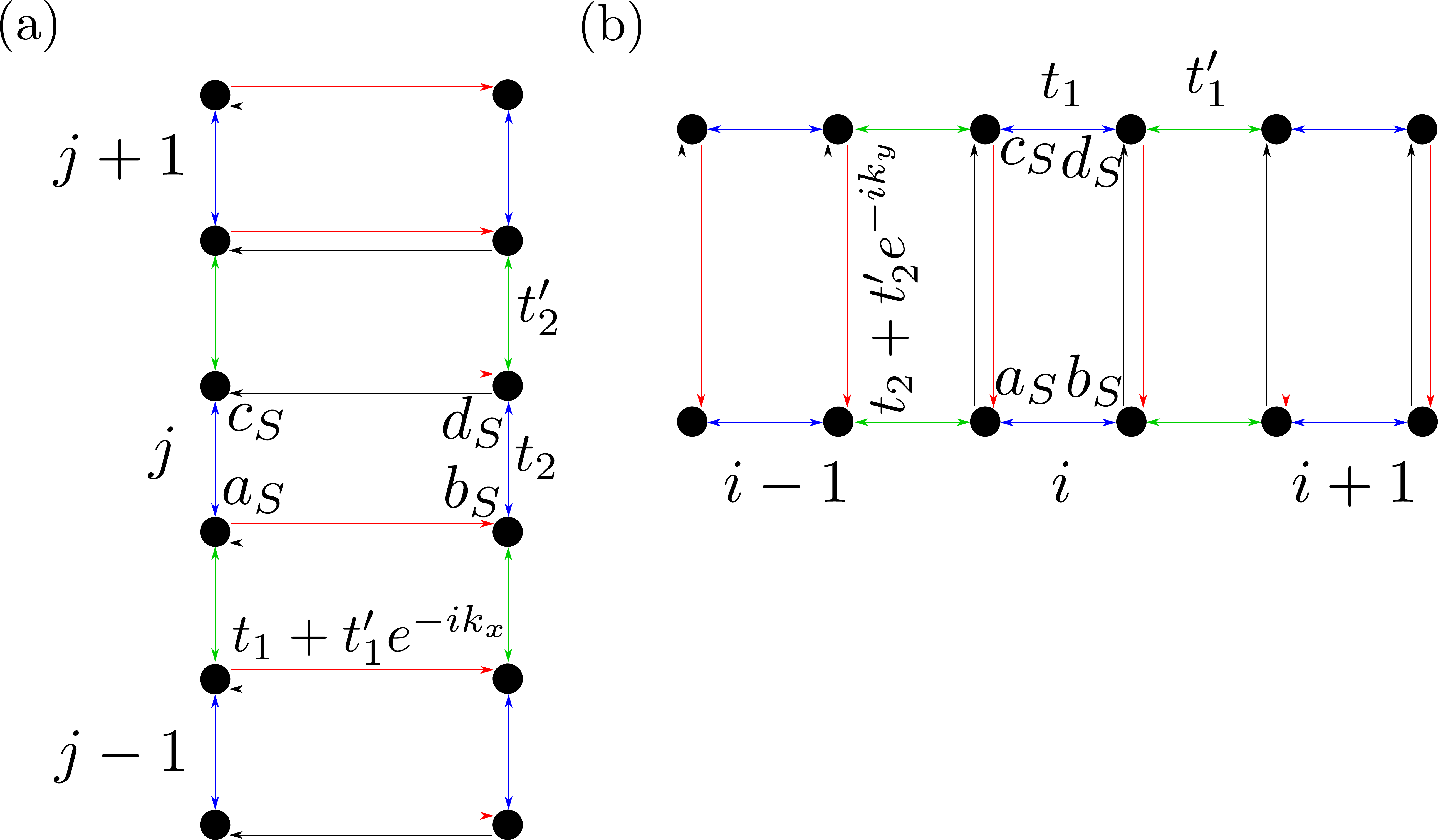}
\caption{(a) Sketch of the vertical 1D model obtained by Fourier-transforming the symmetric lattice Hamiltonian \eqref{Ham_2DSSH_OAM_Sym} along the $x$ direction. (b) Sketch of the horizontal 1D model obtained by Fourier-transforming the symmetric lattice Hamiltonian \eqref{Ham_2DSSH_OAM_Sym} along the $y$ direction. The hopping amplitudes in the directions indicated by black arrows are the complex conjugates of those corresponding to the directions indicated by red arrows.}
\label{1Dmodels_sketch}
\end{figure}

\begin{figure}[t!]
\centering
\includegraphics[width=\linewidth]{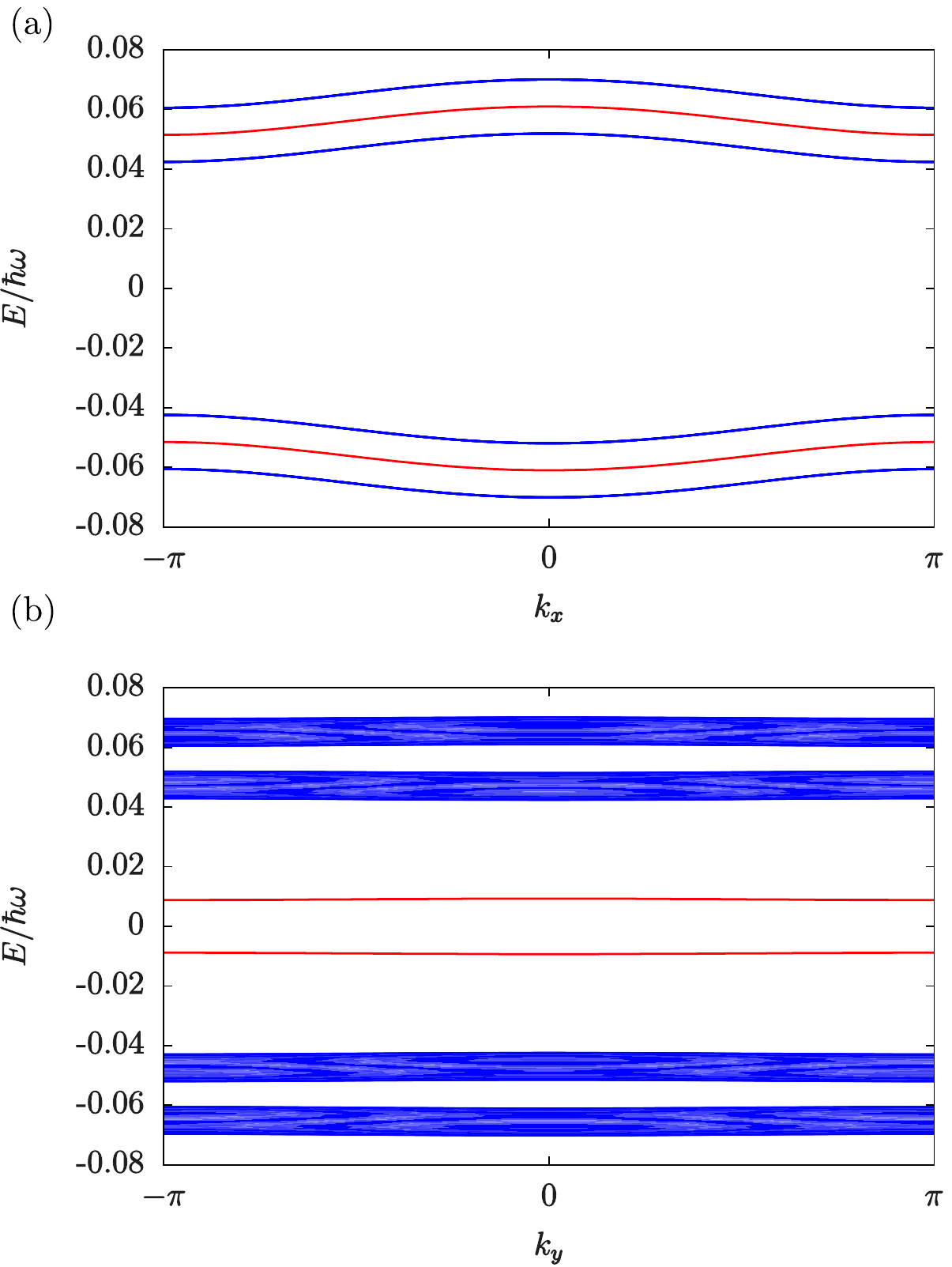}
\caption{Energy spectra of (a) a vertical chain obtained by Fourier-transforming the Hamiltonian of the 2D symmetric lattice along the $x$ direction and (b) a horizontal chain obtained by Fourier-transforming the Hamiltonian of the 2D symmetric lattice along the $y$ direction. In both cases the chains have 50 unit cells and blue (red) curves correspond to bulk (edge) states. 
Each curve of edge states is doubly degenerate. Note that the bulk continua in (b) are much broader than in (a), where bulk states are nearly degenerate at each $k_x$. The parameters of the physical lattice are $R=2.5\sigma$, $s=4\sigma$ and $s'=2\sigma$, for which the coupling parameters of the symmetric lattice are $t_1/t_1'=0.09$, $t_2/t_2'=0.03$.}
\label{1Dmodels}
\end{figure}

For the lattice of symmetric orbitals all the bands have the same values of the 2D Zak's phases. Provided that all the gaps are open, these are 
\begin{equation}
     (\gamma_x^j,\gamma_y^j) = \left\{
	       \begin{array}{ll}
		 (0,0)      & \mathrm{if\ } t_1>t_1',t_2>t_2' \\
		 (\pi,0)      & \mathrm{if\ } t_1<t_1',t_2>t_2' \\
		 (0,\pi)      & \mathrm{if\ } t_1>t_1',t_2<t_2' \\
		 (\pi,\pi)      & \mathrm{if\ } t_1<t_1',t_2<t_2' .\\
	       \end{array}
	       \right.
	     \label{2DZaks}
\end{equation}
Regardless of the values of the coupling parameters, the total polarizations of a non-interacting spinless fermionic system vanish both at half filling (two lower bands occupied) and unit filling (all bands occupied). However, if the $x(y)$ 2D Zak's phases of each band are non-trivial, horizontal (vertical) edge states appear in the energy spectrum of an open lattice. In Fig.~\ref{spectrum_sym_distance} we plot the energy spectrum of a lattice of $10\times 10$ unit cells formed by rings of radius $R=2.5\sigma$ as a function of the inter- and intra-cell separations, keeping their sum constant at the value $s+s'=6\sigma$. For $s<s'$, the couplings fulfill the relations $t_1>t_1',t_2>t_2'$ and no edge or corner states appear in the spectrum. At $s'=s$, the intra- and inter- cell couplings have equal strength and the middle gap closes at zero energy. For $s>s'$, the relations between the couplings are inverted with respect to the case $s<s'$, and horizontal and vertical edge states are present (red curves). The horizontal edge states lie within the gap centered around zero energy and are always detached from the bulk. In contrast, the vertical edge states appear within bulk bands for $3<s/\sigma\lesssim 3.8$. For inter-ring separations larger than $s\simeq 3.8\sigma$, the lower and upper gaps become larger and most of the vertical edge states lie within these gaps [as can also be seen in the energy spectrum for $s=4\sigma$ in Fig.~\ref{SpectrumStates}(a)]. 
The corner states given by the fourfold degenerate black curve in the topological sector are locked to zero-energy. Thus, with the physical system proposed in this paper it is possible to explore the phases $(\gamma_x^j,\gamma_y^j)=(\pi,\pi)$ or $(0,0)$. 
\\
The edge states can also be understood from a different perspective. By Fourier-transforming the Hamiltonian of the lattice of symmetric orbitals \eqref{Ham_2DSSH_OAM_Sym} along only the $y$ ($x$) direction, quasi-one dimensional horizontal (vertical) models with coupling parameters that depend on $k_y$ ($k_x$) are obtained. The Hamiltonians of these models read
\begin{align}
\hat{H}_{S}^{\text{ver}}(k_x)&=\left(t_1+t_1'e^{-ik_x}\right)\sum_{j}
\hat{a}^{j\dagger}\hat{b}_{S}^{j}+
\hat{c}_{S}^{j\dagger}\hat{d}_{S}^{j}
\nonumber\\
&+t_2\sum_{j}
\hat{a}_{S}^{j\dagger}\hat{c}_{S}^{j}+
\hat{b}_{S}^{j\dagger}\hat{d}_{S}^{j}
\nonumber\\
&+t_2'\sum_{j}
\hat{c}_{S}^{j\dagger}\hat{a}_{S}^{j+1}+
\hat{d}_{S}^{j\dagger}\hat{b}_{S}^{j+1}
\nonumber\\
&+\text{H.c.},
\label{Ham_OAM_Sym_1D_Ver}
\end{align}
\begin{align}
\hat{H}_{S}^{\text{hor}}(k_y)&=t_1\sum_{i}
\hat{a}_{S}^{i\dagger}\hat{b}_{S}^{i}+
\hat{c}_{S}^{i\dagger}\hat{d}_{S}^{i}
\nonumber\\
&+t_1'\sum_{i}
\hat{b}_{S}^{i\dagger}\hat{a}_{S}^{i+1}+
\hat{d}_{S}^{i\dagger}\hat{c}_{S}^{i+1}
\nonumber\\
&+\left(t_2+t_2'e^{-ik_y}\right)\sum_{i}
\hat{a}_{S}^{i\dagger}\hat{c}_{S}^{i}+
\hat{b}_{S}^{i\dagger}\hat{d}_{S}^{i}
\nonumber\\
&+\text{H.c.},
\label{Ham_OAM_Sym_1D_Hor}
\end{align}
where the $\hat{p}_{S}^{i}, \hat{p}_{S}^{j}$ operators are respectively the $y$ and $x$ Fourier transforms of $\hat{p}_{S}^{i,j}$. Sketches of the vertical and horizontal 1D models are shown in Fig. \ref{1Dmodels_sketch} (a) and (b) respectively. At $k_x,k_y=0,\pi$, all the couplings in both 1D models become real, allowing to re-express each of them as two decoupled Su-Schrieffer-Heeger (SSH) chains with different on-site potentials. When $\gamma_y^j=\pi$, in the vertical model of \eqref{Ham_OAM_Sym_1D_Ver} four edge states appear in two-fold degenerate pairs, as exemplified in Fig. \ref{1Dmodels} (a) (red lines). Since this model has been obtained from assuming periodic boundary conditions in the $x$ direction of the original lattice, these 1D edge states correspond to vertical edge states in the 2D model. Similarly, as exemplified in Fig. \ref{1Dmodels} (b), when $\gamma_x^j=\pi$ the 1D horizontal model has edge states that correspond to horizontal edge states in the original 2D lattice. In the following section, we will discuss how the analysis of the bands of edge states of the 1D models can be used as a way to characterize the second-order topological properties of the model.
\subsection{Second-order topological effects and corner states}
\begin{figure}[t!]
\centering
\includegraphics[width=\linewidth]{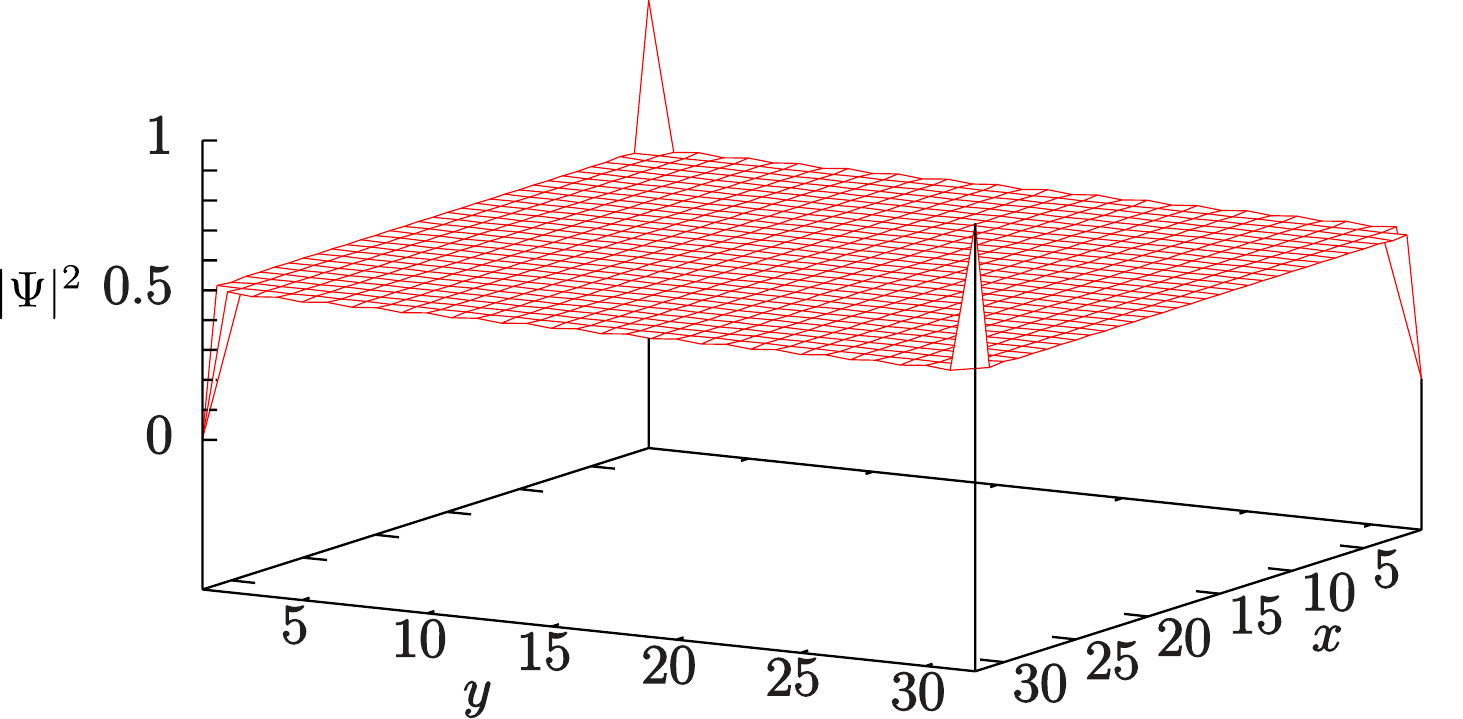}
\caption{Atomic density distribution of a symmetric lattice of $16 \times 16$ unit cells for a non-interacting spinless fermionic gas at half filling. The parameters of the physical lattice are $R=2.5\sigma$, $s=4\sigma$ and $s'=2\sigma$, for which the coupling parameters of the symmetric lattice are $t_1/t_1'=0.09$, $t_2/t_2'=0.03$, $t_2'/t_1'=-0.16$. A small perturbation that breaks the chiral symmetry has been introduced in the numerical calculations.}
\label{CharHalfFill}
\end{figure}

In the $(\gamma_x^j,\gamma_y^j)=(\pi,\pi)$ phase, a finite lattice has four zero-energy corner states. By introducing a small perturbation that breaks chiral symmetry, this degeneracy is lifted and the states become localized at specific corners. In this situation, at half filling of the symmetric lattice (which, due to the degeneracy between the symmetric and antisymmetric orbitals, corresponds to unit filling of the physical sites of the original model with two OAM states per ring), only two of the corner states are populated and the total density distribution, defined as the sum of the densities of all the occupied states, has bright and dark peaks at the corners, as shown in Fig. \ref{CharHalfFill}. These density peaks are analogous to charge concentrations in an electronic system, and thus give rise to the atomic analogs of the edge polarizations and quadrupole moment.

Recently, a successful method to characterize the topological quadrupole moment has been proposed \cite{bernevigPRB}. In order for a finite topological quadrupole moment to arise, at least two bands have to be occupied at half filling, as is the case in our model. Nevertheless, a necessary condition for the procedure to work is that the $x$ and $y$ mirror symmetries do not commute. In the minimal model for a bulk quadrupole insulator studied in \cite{bernevigPRB}, this non-commutativity between the reflection symmetries is achieved by introducing a $\pi$ flux in each plaquette through alternating signs in the vertical couplings. In our model, however, the inversion symmetries commute and the quadrupole moment can not be characterized following the recipe presented in \cite{bernevigPRB}. One way to circumvent this difficulty is to redefine the quadrupole moment per unit area as 
\begin{eqnarray}
q_{xy}&=&\frac{1}{2}\tilde{P}_x\tilde{P}_y,
\label{quadmom}
\\
\tilde{P}_{x(y)}&=&\sum_{j\in O}P_{x(y)}^j,
\end{eqnarray}
where $O$ defines the set of occupied bands ($O=\{1,2\}$ in our case) and $\tilde{P}_{x(y)}$ is the direct sum of the polarizations along the $x(y)$ direction.
Eq. \eqref{quadmom} implies that the bulk quadrupole moment is $q_{xy}=\frac{1}{2}$ in the $(\gamma_x^j,\gamma_y^j)=(\pi,\pi)$ phase, that is, when both vertical and horizontal edge states are present, and $q_{xy}=0$ otherwise. This is in accordance with our numerical calculations and allows to regard $q_{xy}$ as the topological index associated to the appearance of corner states. The polarization $P_{x(y)}$ at half filling is related to $\tilde{P}_{x(y)}$ as $P_{x(y)}=\tilde{P}_{x(y)} \mod{1}$. In our model we have $P_{x(y)}=0$ for both $\tilde{P}_{x(y)}=0$ (trivial phases) and $\tilde{P}_{x(y)}=1$ (non-trivial phase), that is, the system is not polarized at half filling in either direction, which is why $P_{x(y)}$ cannot be used in the definition of $q_{xy}$, since it is insensitive to transitions between different second-order topological regimes.
\begin{table}[t!]
\begin{tabular}{|c|c|c|c|c|}
\hline
 $(\gamma_x^j,\gamma_y^j)$ &  \tiny{Horizontal edge states } &  \tiny{Vertical edge states} &  \tiny{Corner states}&$q_{xy}$\\
 \hline
 $(0,0)$ & No& No& No& 0\\
 \hline
  $(\pi,0)$ & Yes, $\gamma_{\text{edge}}^x=0$& No& No& 0\\
 \hline
  $(0,\pi)$ & No& Yes, $\gamma_{\text{edge}}^y=0$& No& 0\\
 \hline
  $(\pi,\pi)$ & Yes, $\gamma_{\text{edge}}^x=\pi$& Yes, $\gamma_{\text{edge}}^y=\pi$& Yes& $\frac{1}{2}$\\
 \hline
\end{tabular}
\caption{Possible combinations of values of the 2D Zak's phases of the bulk and the Zak's phases of the bands of edge states of the 1D horizontal and vertical models. The last two columns indicate the presence of corner states in the 2D lattice under open boundary conditions in both $x$ and $y$ and the values for the quadrupole moment defined in \eqref{quadmom}.}
\label{TableZaks}
\end{table}
Alternatively, the presence of edge polarizations and a finite quadrupole moment can also be tested by analysing the 1D models that are obtained by Fourier-transforming the 2D lattice along the $x$ ($y$) direction. As we discussed in Sec. \ref{TopProp} A, the horizontal and vertical edge states can also be seen as edge states of these 1D models. If one considers chains with finite sizes in the $y$ ($x$) direction and periodic boundaries in the $x$ ($y$) direction and diagonalizes them as a function of $k_x$ ($k_y$), four bands of edge states (coming in two-fold degenerate pairs, see Fig.~\ref{1Dmodels}) are obtained if the original 2D lattice has bands with non-trivial 2D Zak's phases in the $x$ ($y$) axis. 
The degeneracy of the edge bands can be lifted by introducing a small perturbation that breaks the chiral symmetry, allowing to compute their Zak's phases, $\gamma_{\text{edge}}^x$ and $\gamma_{\text{edge}}^y$. 
A non-trivial Zak's phase in the edge states indicates the presence of ``edge of edge" states (i.e. corner states) and a finite quadrupole moment at half filling for a non-interacting spinless fermionic gas. 
The topological behavior of our model as a function of the values of the 2D Zak's phases in \eqref{2DZakX} and \eqref{2DZakY} is summarized in Table \ref{TableZaks}. 
The only topological phase of the bulk in which the edge bands have non-trivial Zak's phases is  $(\gamma_x^j,\gamma_y^j)=(\pi,\pi)$. Thus, the simultaneous non-triviality of the Zak's phases of edge bands of the 1D models $\gamma_{\text{edge}}^x,\gamma_{\text{edge}}^y$ is in one to one correspondence with the appearance of corner states and a finite quadrupole moment in an open 2D lattice. More specifically, each pair of symmetric edge bands with a non-trival 1D Zak's phase has two zero-energy corner-states associated with it. Each of these states is shared by a vertical and a horizontal edge band. Therefore, the four corner states that appear at zero energy are associated with the four occupied bands of edge states at half filling, of which two correspond to the 1D horizontal model and two to the 1D vertical model.
\subsubsection*{Symmetry protection of the corner states}
\begin{figure}[t!]
\centering
\includegraphics[width=\linewidth]{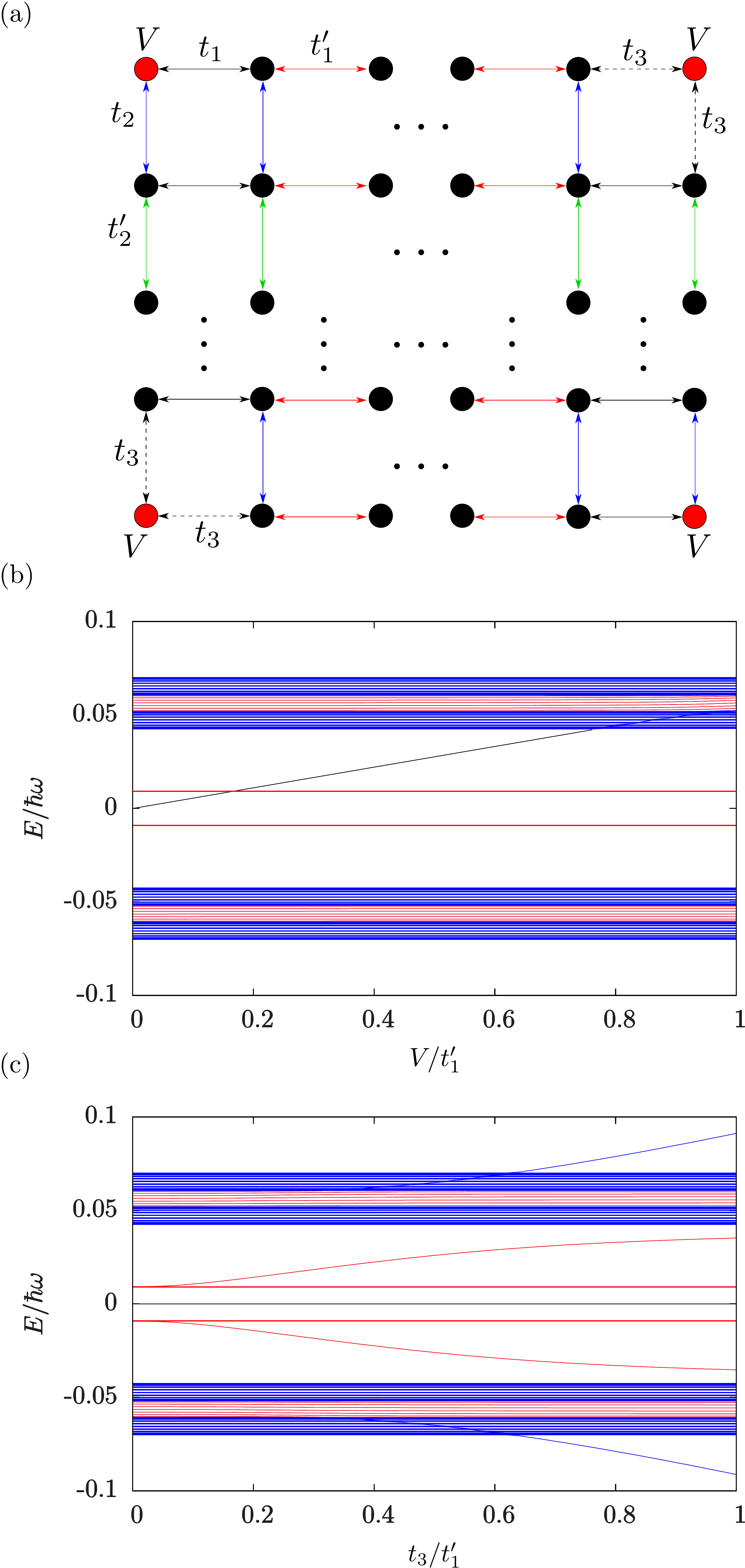}
\caption{(a) Sketch of the different types of perturbations described in the main text: on-site potential V at the corners, which preserves the reflection symmetries but not the chiral one, and modified $t_3$ coupling in two of the corners, which has an effect opposite to $V$. (b) Spectrum of a lattice of ${10\times 10}$ unit cells as a function of $V$ leaving the corner couplings unchanged. (c) Spectrum of the same lattice as in (b) as a function of $t_3$ keeping $V=0$. The parameters of the physical lattice are $R=2.5\sigma$, $s=4\sigma$ and $s'=2\sigma$, for which the coupling parameters of the symmetric lattice are $t_1/t_1'=0.09$, $t_2/t_2'=0.03$, $t_2'/t_1'=-0.16$.}
\label{perturbations}
\end{figure}
Before concluding, let us briefly discuss the symmetries that are responsible for the topological protection of the corner states. While the quantization of the bulk polarizations $P_x, P_y$ and the quadrupole moment $q_{xy}$ is ensured by the $x$ and $y$ mirror symmetries, it is the chiral symmetry of $\hat{H}_S$ that protects the corner states. This can be justified by taking into account the fact that the spectrum of an Hamiltonian is symmetric around zero-energy in the presence of chiral-symmetry, implying that the zero-energy corner modes are eigenstates of the chiral operator \cite{CourseTopology}. Therefore, the corner states are not affected by perturbations that preserve chiral-symmetry.

In Fig. \ref{perturbations} (a) we illustrate two different kinds of perturbations. On the one hand, we consider an on-site potential $V$ acting only on the corners of the lattice, which preserves the $x$ and $y$ reflection symmetries but breaks the chiral symmetry. On the other hand, we substitute in two of the corners the couplings of the model by a different one named $t_3$. This perturbation has an opposite effect to $V$, i.e., it breaks the reflection symmetries but preserves the chiral one. In Fig. \ref{perturbations} (b) we plot the spectrum of a finite lattice as a function of $V$ leaving the corner couplings unchanged. As $V$ increases, the energy of the corner modes (black line) increases until they merge into the bulk. Fig \ref{perturbations} (c) shows the spectrum of the same lattice as in Fig \ref{perturbations} (b) but for $V=0$ and increasing $t_3$ until it reaches the value $t_3=t_1'$, which is the largest coupling of the symmetric lattice. Since this perturbation preserves chiral-symmetry, all corner states, including the two localized around the corners with perturbed edge couplings $t_3$, remain locked at zero energy regardless of the value of $t_3$.
\section{Conclusions}
\label{conclusions}
We have shown that ultracold atoms carrying OAM in arrays of cylindrically symmetric potentials can be used to implement a second-order topological insulator with zero-energy corner states and a quantized quadrupole moment. The topological properties of the system can be more directly analyzed by performing a change of basis that transforms the original model with two states per site into two independent lattices with one $p_x-$ or $p_y-$like orbital per site, formed respectively by symmetric and antisymmetric combinations of the OAM states. We have shown that for experimentally realistic parameters the system can display zero-energy corner modes associated with the non-trivial second-order topological phase. A new expression for the quantized quadrupole moment, involving a new quantity defined here as the direct sum of the polarizations of the occupied bands, was identified here as the relevant second-order topological index. A complementary approach to the topological characterization of the corner states, related to an analysis of the Zak's phases of the edge bands that appear when periodic boundary conditions are imposed alternately along one of the directions, while keeping the other open, is shown to be consistent with the former.

In an experimental implementation, the edge and corner states could be prepared by first populating only the corresponding sites of the lattice in the limit of zero intra-cell couplings and then adiabatically turning them on, as done in \cite{SSHmom} to obtain the edge states of the SSH model in a system of ultracold atoms. The half-filled state, in which the quantized quadrupole moment is manifested through the density distribution, could be realized using a gas of spin-polarized fermions formed by as many atoms as sites in the lattice, in such a way that the states with energy below the Fermi level would be consecutively occupied. In order to image these states, a quantum gas microscope, which provides real-space images with single-site resolution \cite{QGasMic}, could be employed. We note that the topological edge states of the SSH model have been imaged in systems of ultracold atoms in optical lattices both in momentum \cite{SSHmom} and in real \cite{SSHreal} space.
 
As a final remark, we note that the model studied in this paper could be implemented with other systems that support OAM modes or $p_x$ and $p_y$ orbitals, such as artificial electronic lattices \cite{pBandElectronic}, photonic waveguides \cite{reviewLonghi,OAMwaveguides} or polariton resonators, where the edge states of the SSH model have already been observed \cite{SSHPolaritons}.
\acknowledgements
G.P., J.M., and V.A. gratefully acknowledge financial support  from  the  Ministerio  de  Econom\'ia  y  Competitividad, MINECO, (FIS2017-86530-P)  and  from the Generalitat de Catalunya (SGR2017-1646). G.P. acknowledges  financial  support  from  MINECO  through  Grant  No. BES-2015-073772 and a travel grant from the COST Action CA16221.  A.M.M.  acknowledges  financial  support from the Portuguese Institute for Nanostructures, Nanomodelling  and  Nanofabrication  (I3N)  through  the  Grant  No. BI/UI96/6376/2018. A.M.M. and R.G.D. acknowledge funding by the FEDER funds through the COMPETE 2020 Programme and National Funds through FCT-Portuguese Foundation  for  Science  and  Technology  under  the  Project  No. UID/CTM/50025/2013 and under the Project No. PTDC/FIS-MAC/29291/2017.

\end{document}